\documentclass[aps,prl,twocolumn,superscriptaddress,amsmath,amssymb,floatfix]{revtex4-2}
\usepackage{graphicx}
\usepackage{xcolor}
\usepackage[colorlinks=true,linkcolor=blue,citecolor=blue,urlcolor=blue]{hyperref}

\newcommand{\NVm}{\ensuremath{\mathrm{NV}^{-}}}
\newcommand{\NVn}{\ensuremath{\mathrm{NV}^{0}}}
\newcommand{\tA}{{}^{3}A_{2}}
\newcommand{\tE}{{}^{3}E}
\newcommand{\qA}{{}^{4}A_{2}}
\newcommand{\dE}{{}^{2}E}
\newcommand{\eV}{\,\mathrm{eV}}

\begin{document}

\title{Quartet state of the neutral nitrogen--vacancy center in diamond and optimized spin readout under
near-infrared two-photon excitation}

\author{Nikoletta Jegenyes}
\affiliation{HUN-REN Wigner Research Centre for Physics, P.O.\ Box 49, H-1525 Budapest, Hungary}
\author{P\'al Ma\'ak}
\affiliation{Department of Atomic Physics, Institute of Physics, Budapest University of
Technology and Economics, M\H{u}egyetem rakpart 3., H-1111 Budapest, Hungary}
\affiliation{HUN-REN Wigner Research Centre for Physics, P.O.\ Box 49, H-1525 Budapest, Hungary}
\author{Vladimir Verkhovlyuk}
\affiliation{HUN-REN Wigner Research Centre for Physics, P.O.\ Box 49, H-1525 Budapest, Hungary}
\author{L\'aszl\'o Himics}
\affiliation{HUN-REN Wigner Research Centre for Physics, P.O.\ Box 49, H-1525 Budapest, Hungary}
\author{Tam\'as V\'aczi}
\affiliation{HUN-REN Wigner Research Centre for Physics, P.O.\ Box 49, H-1525 Budapest, Hungary}
\author{Mikl\'os Veres}
\email{veres.miklos@wigner.hun-ren.hu}
\affiliation{HUN-REN Wigner Research Centre for Physics, P.O.\ Box 49, H-1525 Budapest, Hungary}
\author{\'Ad\'am Gali}
\email{gali.adam@wigner.hun-ren.hu}
\affiliation{HUN-REN Wigner Research Centre for Physics, P.O.\ Box 49, H-1525 Budapest, Hungary}
\affiliation{Department of Atomic Physics, Institute of Physics, Budapest University of
Technology and Economics, M\H{u}egyetem rakpart 3., H-1111 Budapest, Hungary}
\affiliation{MTA-WFK Lend\"ulet ``Momentum'' Semiconductor Nanostructures Research Group,
P.O.\ Box 49, H-1525 Budapest, Hungary}

\date{\today}

\begin{abstract}
Photoionization of the excited nitrogen--vacancy center (NV$^-$) in diamond leaves the neutral defect in a
metastable quartet state of unknown energy. Using two-photon optically detected magnetic resonance with a
tunable femtosecond laser, we resolve its onset in the spin contrast at $1.23\pm0.02$~eV and, with the
ground-state threshold from the same setup, place the quartet $0.54\pm0.04$~eV above the neutral ground
state. The threshold sets the optimum excitation for spin readout at 1060--1075~nm, in the near-infrared
tissue window, with 26--37\% higher shot-noise-limited sensitivity than at 1030--1040~nm.
\end{abstract}

\maketitle

\begin{figure*}[t]
\centering
\includegraphics[width=\textwidth]{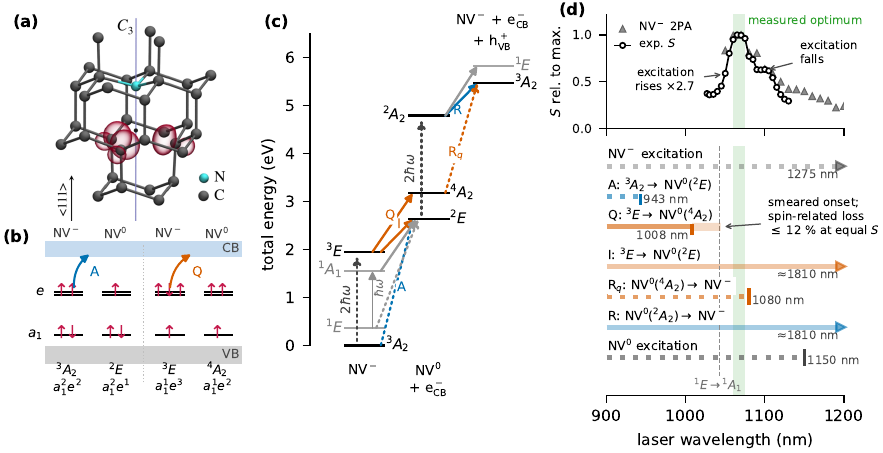}
\caption{(a) NV center: nitrogen (N) next to a vacancy (dot), $C_3$ axis along $\langle111\rangle$, and
spin density of \NVm{} on the three carbon dangling bonds (red); adapted from Ref.~\cite{Bockstedte2018}.
(b) One-electron $a_1$ and $e$ levels in the gap and their occupation; A removes an $e$ electron from
$\tA$ ($\to\dE$), Q the minority-spin $e$ electron from $\tE$, leaving three parallel spins ($\to\qA$). (c) Total-energy diagram of \NVm{} and \NVn{} + $e^-_{\rm CB}$ with the excitation
($2\hbar\omega$, dotted gray), ionization (A, I, Q) and recharge (R, R$_q$) channels; orange,
spin-related; blue, spin-blind; dotted, two-photon steps; gray, channels through the singlets
(ionization of $^1A_1$ and $^1E$, recharge into $^1E$), which leave no separate signature (see
text). The singlets are drawn at an illustrative singlet--triplet gap of 0.36~eV. (d) Top: measured photoluminescence $S$
(300~mW, circles) and the \NVm{} two-photon absorption cross section of Ref.~\cite{Ji2018}
(triangles), each normalized at its maximum. Bottom: the channels of (c) in the order of the cycle, each drawn on the
open side of its threshold (vertical tick) for the number of photons it needs; \NVm{} excitation, I and R are open over the whole range
(arrows, with the threshold beyond it); the singlet channels, whose thresholds depend on the singlet--triplet gap, are omitted; the shaded part of the Q bar marks the onset broadened by the inhomogeneous width and the laser bandwidth. Dashed line,
$^1E\to{}^1A_1$ resonance (1042~nm); green band, measured optimum.}
\label{fig:concept}
\end{figure*}

\paragraph*{Introduction.}
The negatively charged nitrogen--vacancy center (\NVm) in diamond, a substitutional nitrogen next to
a vacancy [Fig.~\ref{fig:concept}(a)], is the leading solid-state
quantum sensor~\cite{Doherty2013,Schirhagl2014,Gali2019,Barry2020}. Two-photon excitation in the near
infrared adds intrinsic optical sectioning and deeper penetration into scattering tissue; it has
been used to image fluorescent nanodiamonds in cells~\cite{Wee2007,Chang2008,Ju2022} and, recently,
to read out the spin by optically detected magnetic resonance (ODMR)~\cite{Nguyen2026}. Under
intense femtosecond excitation, however, the defect is ionized~\cite{Su2025}, and the spin contrast depends on
which ionization channels the laser opens.

The central channel is photoionization of the spin-polarizing excited state $\tE$. Ejecting an
$e$ electron from the $a_1^1e^3$ configuration of $\tE$ leaves \NVn{} in $a_1^1e^2$, whose lowest
state is the quartet $\qA$, not the $a_1^2e^1$ ground state $\dE$~\cite{Razinkovas2021}
[Fig.~\ref{fig:concept}(b)]. The existence of the $\qA$ quartet state has been established by electron
paramagnetic resonance (EPR) measurements~\cite{Felton2008,Gali2009}, but its energy relative to the $\dE$ ground
state remains uncertain: calculations
place it 0.48--0.68~eV above $\dE$~\cite{ThieringGali2024,Pfaffle2021}, an estimate from room-temperature
photoionization thresholds gives $\approx0.4\eV$~\cite{ThieringGali2024}, and a cavity experiment
at two fixed infrared wavelengths bounds it by $\leq0.58\eV$~\cite{Shandilya2024}. One-photon
ionization of $\tE$ has otherwise been used at a single wavelength, 1064~nm~\cite{Zhang2021}.

In this Letter we resolve the onset of the $\tE\to\NVn(\qA)$ channel spectrally. With one tunable
source we measure its threshold $E_q$ and the ionization potential of the ground state,
$\mathrm{IP}(\tA)$, on the same sample at the same temperature, which gives the quartet energy
without assumptions about temperature shifts, and locate the excitation wavelength that maximizes
the sensitivity of two-photon ODMR of NV$^-$.

\paragraph*{Channels and thresholds.}
Figure~\ref{fig:concept}(c) shows the channels in a total-energy diagram. Two photons excite
\NVm{} from $\tA$ to $\tE$ (zero-phonon line $E_{\rm ZPL}=1.945\eV$). Two photons ionize the ground state into
$\NVn(\dE)$ above $\mathrm{IP}(\tA)$ (channel~A); this route is spin-blind. A third photon ionizes
the freshly excited $\tE$, either into $\qA$ (Q) or into $\dE$ (I); both act on a
spin-polarized state and remove spin contrast. Excited \NVn{} is recharged by one further photon
(R), the quartet by two (R$_q$). The thresholds are linked by~\cite{Razinkovas2021}
\begin{equation}
E_q=\mathrm{IP}(\tA)-E_{\rm ZPL}+\left[E(\qA)-E(\dE)\right].
\label{eq:cycle}
\end{equation}
By Eq.~(\ref{eq:cycle}), an onset near 1.2~eV requires a final state about 0.5~eV above $\dE$. Group
theory and Hund's rule leave a single candidate: the $\qA$ quartet, the state identified by its EPR
signal~\cite{Felton2008,Gali2009}. Channel I is allowed by direct ionization into the $a_1^1e^2$ admixture
(8--28\%) of the correlated $\dE$ ground state~\cite{ThieringGali2024}; an Auger
process~\cite{Siyushev2013} may contribute.

Figure~\ref{fig:concept}(d) places the thresholds on the laser axis together with the measured
excitation spectrum. The optimum is expected between the last spin-destroying onset (Q, R$_q$) and
the fall of the two-photon absorption of \NVm{}~\cite{Ji2018}.

\paragraph*{Experiment.}
The sample is a single-crystal $\langle100\rangle$ diamond grown by chemical vapor deposition,
from Diamond Elements Pvt.\ Ltd.~\cite{DiamondElements} ([N]~$=10$~ppm, 0.3~mm thick), with an NV
density estimated at $\approx0.1$~ppm ($\mathrm{NV/N}\approx1\%$); it is measured in a
single-objective epi-configuration. The beam of a tunable Coherent Chameleon Discovery
femtosecond laser (680--1300~nm, 80~MHz, 100--150~fs, 13--16~nm spectral width; wavelength
measured with a high-resolution spectrometer, APE waveScan) is attenuated by a half-wave plate and a
polarizing beamsplitter cube, transmitted through an 805~nm longpass dichroic mirror and focused by an
Olympus Plan~N $40\times$/0.65 objective. The photoluminescence returns through the objective, is
reflected by the dichroic mirror and
integrated over 690--750~nm on an avalanche photodiode. Continuous microwaves (15~dBm),
modulated at 50~Hz by the lock-in reference, are delivered by a wire loop; details are given in the
Supplemental Material (SM)~\cite{SM}. At each wavelength
the photoluminescence $S$ (minus the
laser-blocked dark level) and the spin signal $\Delta S$ (the decrease of $S$ on resonance) are read at the same setting, one
immediately after the other; the contrast is $C=\Delta S/S$. All data are taken at room
temperature.

\begin{figure}[t]
\centering
\includegraphics[width=\columnwidth]{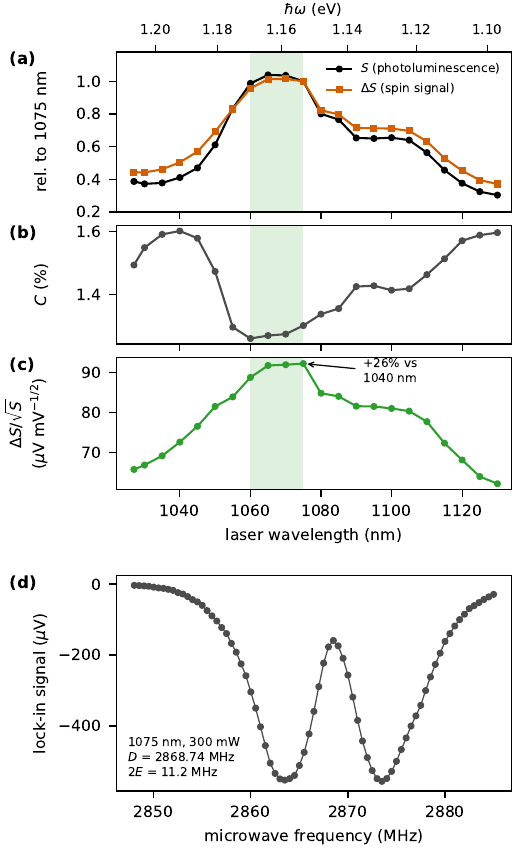}
\caption{Wavelength scan at 300~mW. (a) Photoluminescence $S$ and spin signal $\Delta S$, relative to
1075~nm; top axis, photon energy. (b) Contrast $C=\Delta S/S$. (c) Shot-noise figure of merit
$\Delta S/\sqrt{S}$. Green band, 1060--1075~nm. (d) ODMR
spectrum at 1075~nm (lock-in signal, in phase; the photoluminescence decreases on resonance).}
\label{fig:observation}
\end{figure}

Figure~\ref{fig:observation} shows a wavelength scan at 300~mW across 1027--1130~nm. $S$ and $\Delta S$
peak together at 1060--1075~nm (a). The contrast is lowest there (b): 1.27--1.30\% against
1.55--1.60\% at 1030--1040~nm. The shot-noise figure of merit $\Delta S/\sqrt{S}=C\sqrt{S}$,
however, peaks in the same window (c), at 91.7--92.2 against 72.6 at 1040~nm (+26\%) and 66.8 at
1030~nm (+37\%). The resonances are those of \NVm{} [(d); $D=2868.74$~MHz, $2E=11.2$~MHz].

\begin{figure}[t]
\centering
\includegraphics[width=\columnwidth]{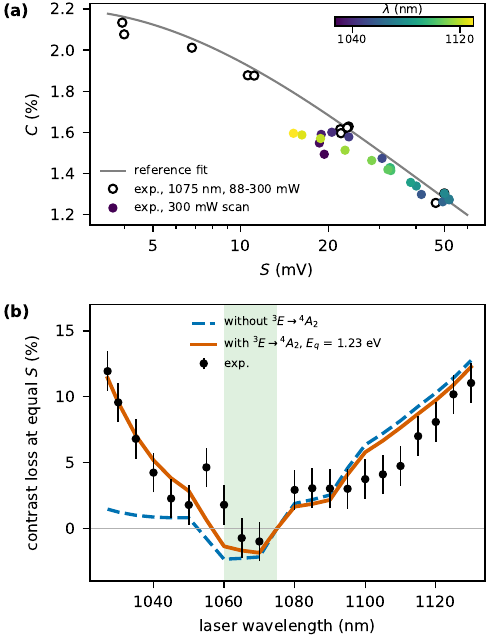}
\caption{(a) Contrast against photoluminescence: 1075~nm at 88--300~mW (open circles) with the
reference fit, and the 300~mW scan colored by wavelength. (b) Contrast loss at equal $S$ for the
300~mW scan (error bars, 1.5\%), with the rate model fitted to all 52 values, without (dashed)
and with (solid, $E_q=1.23\eV$) the $\tE\to\NVn(\qA)$ channel; $\sigma_I=0.13\,\sigma_Q$.}
\label{fig:physics}
\end{figure}

\paragraph*{Spin contrast at equal photoluminescence.}
The raw contrast mixes two effects: $C$ falls as $S$ rises, and both change with wavelength. We
therefore compare each point with the contrast measured at 1075~nm at the same $S$, by varying the
power from 88 to 300~mW [Fig.~\ref{fig:physics}(a)], and define the contrast loss at equal photoluminescence
$L=1-C/C_{1075}(S)$. A loss channel that is absent at 1075~nm appears as $L>0$.

\paragraph*{Rate model.}
To relate the contrast loss to the channels of Fig.~\ref{fig:concept} we use a rate model of an NV
center in both charge states, \NVm{} and \NVn{} [Fig.~\ref{fig:concept}(c)]. It follows the populations of ten levels: the $m_s=0$ and $\pm1$
sublevels of $\tA$ and $\tE$, the singlets $^1A_1$ and $^1E$, the $\dE$ and $^2A_2$ states of
\NVn{}, and the $m_s=\pm1/2$ and $\pm3/2$ sublevels of $\qA$. Each laser pulse is much shorter than
any decay and is treated as instantaneous: it transfers population with probabilities set by the
photon fluence and the cross sections of the two-photon excitations and of the ionization and
recharge channels, including the ionization of $\tE$ excited by the same pulse; the lowest triplet
above $\tE$ lies 2.385~eV higher~\cite{Luu2025}, so $\tE$ absorbs our photons only into the
continuum. The cross sections
depend on the photon energy through their zero-phonon thresholds, broadened by the calculated phonon
sidebands~\cite{Razinkovas2021}, an inhomogeneous width and the laser spectrum. Between pulses the populations follow linear rate equations for radiative decay,
intersystem crossing, the decay of the singlets and of the quartet, spin--lattice relaxation, the
microwave drive, electron transfer from $\tE$ to nearby nitrogen donors, and the return of
\NVn{} to \NVm{} by electron tunnelling from the donors, with the distribution of tunnelling
rates set by the random distance to the nearest donor at $[\mathrm{N}]=10$~ppm~\cite{Alkauskas2025}. 
The steady state of the 80~MHz pulse train is the fixed point of the map over one period, found numerically. The
photoluminescence follows from the radiative decays of $\tE$ and $^2A_2$, weighted by their
emission into the detection window; $S$ and $C$ are obtained with and without microwaves and
averaged over the intensity profile of the focus. The rate equations, the rates and cross sections with their sources, and the numerical methods
are given in Ref.~\cite{SM}.

\begin{figure}[t]
\centering
\includegraphics[width=\columnwidth]{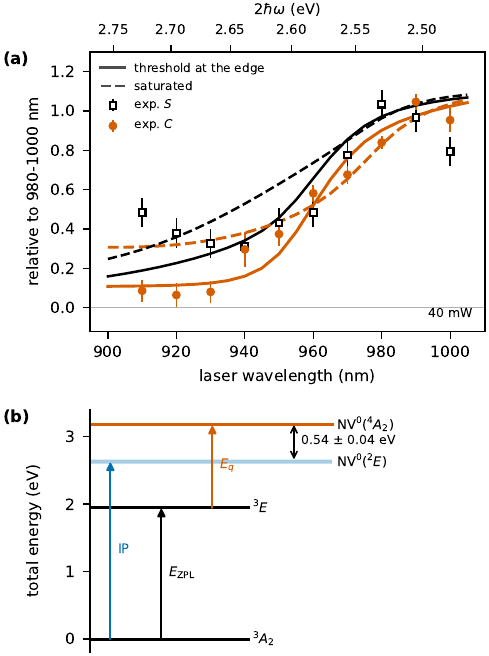}
\caption{(a) Scan at 40~mW: $S$ (squares) and $C$ (circles) relative to 980--1000~nm. Lines: model $S$ (black)
and $C$ (orange) from the fit to both together with the contrast-loss values (solid, $\mathrm{IP}=2.63\eV$), and for a
saturated ground-state channel (dashed, $\mathrm{IP}=2.54\eV$); top axis, two-photon energy. (b) Energy ladder of Eq.~(\ref{eq:cycle}): $\mathrm{IP}(\tA)$ from
(a), $E_q$ from Fig.~\ref{fig:physics} and $E_{\rm ZPL}$ give the quartet energy.}
\label{fig:threshold}
\end{figure}

The contrast loss is zero within 1\% at 1060--1075~nm and rises on both sides
[Fig.~\ref{fig:physics}(b)]. We fit the model to 52 contrast-loss values measured at
150--300~mW~\cite{SM}; each value carries an uncertainty of 1.5\%, set by the choice
of the reference measurements. The free parameters are the strengths of channels A, Q and R$_q$, the
two-photon absorption of \NVn{} and the \NVn{} share of the detected emission, with
$\mathrm{IP}(\tA)$ fixed at the value obtained together with the scan of Fig.~\ref{fig:threshold}, and $E_q$ scanned. 
With channel Q, and with the cross section of channel I tied to that of Q, $\sigma_I=r\sigma_Q$,
the model reproduces these values with a root-mean-square deviation of
1.4\%, within that uncertainty, for $E_q=1.23\eV$ [Fig.~\ref{fig:physics}(b)]. Without channel Q the deviation is
9.1\%, up to 29\% at the shortest wavelengths. The visible onset near 1055~nm is the tail of
this zero-phonon threshold, broadened by the phonon sideband, an inhomogeneous width of 20~meV and the laser
bandwidth. Inhomogeneous widths of 10--40~meV, ratios $r=0.08$--0.13 ($r=0.28$ fits worse) and a shift of
IP by $\pm0.03$~eV move the best threshold within 1.215--1.245~eV; together this gives $E_q=1.23\pm0.02\eV$. The
fits need a photo-recharge of $\qA$ (R$_q$), which we take at least as strong as channel A, as calculated for the
corresponding one-photon transitions~\cite{Bourgeois2017}; $\qA$ then stays below 0.4\% populated, as assumed for
green excitation~\cite{Wirtitsch2023}.

The long-wavelength side carries independent information. Beyond 1080~nm the two-photon absorption
falls, and matching $S$ requires a higher photon flux per excitation, so any one-photon ionization
of $\tE$ grows relative to 1075~nm. Channel Q is closed there. Without channel I the model gives
$L\approx0$ at 1080--1130~nm, against 3--11\% measured; the root-mean-square deviation of these
11 values grows from 1.7\% to 6.9\%, and that of all 52 values from 1.4\% to 3.4\%. The data
thus require a $\tE\to\NVn(\dE)$ channel with $\sigma_I=(1.3$--$2.3)\times10^{-19}$~cm$^2$, an
order of magnitude below direct ionization into $\qA$ at equal excess energy but not negligible: it
ionizes 11--22\% of the excited $\tE$ per pulse. Beyond 1030~nm, where only the broadened tail of the Q onset
remains, this is several times the fraction ionized into the quartet, so that channel I is the main
ionization route of $\tE$ there. The correlated $\dE$ well
accounts for it. A cavity experiment likewise requires $\tE\to\dE$ ionization at 1524~nm, below the
quartet threshold, about an order of magnitude weaker than $\tE\to\qA$~\cite{Shandilya2024}.

\paragraph*{Channels of the singlets.}
The singlet shelf of \NVm{} adds three channels in the same spectral range: ionization of the
short-lived $^1A_1$~\cite{UlbrichtLoh2018} after the $^1E\to{}^1A_1$ resonance at
1042~nm~\cite{Rogers2008,Acosta2010,Kehayias2013,Younesi2025}; two-photon ionization of $^1E$, whose one-photon
threshold has been measured with visible light~\cite{Wolf2023,Blakley2024,Ung2025}; and recharge of excited \NVn{}
into the singlets instead of $\tA$, whose threshold is known from transient absorption~\cite{LuuMQT2024}. Their thresholds
depend on the singlet--triplet gap and overlap with those of Q, R$_q$ and I, and several of
these processes act in parallel with partly opposite effects on the spin contrast. None of them
leaves a separate signature within the 1.5\% uncertainty, as three tests show~\cite{SM}.
(i) Ionization of $^1A_1$ would make the contrast loss at the 1042~nm resonance grow with power.
We measure a growth of at most 6\% ($2\sigma$) from 150 to 300~mW, as expected without this route ($+2.5\%$); this bounds the route to about a
third or less of an estimate from the measured $^1E\to{}^1A_1$ absorption, the selection rules and the calculated
photoionization of $\tA$~\cite{Razinkovas2021}. (ii) Our photons
(0.99--1.36~eV) lie below the one-photon threshold of $^1E$, so $^1E$ is ionized only by
nonresonant two-photon absorption; its cross section is below about 1~GM
(G\"oppert-Mayer unit, $1~\mathrm{GM}=10^{-50}$~cm$^4$\,s; see SM~\cite{SM}), comparable to the two-photon absorption of \NVm{}~\cite{Wee2007,Ji2018}.
(iii) Recharge into the singlets does not improve the fit. These channels are therefore omitted
from the fits shown here.

The photoluminescence step below 1060~nm [Fig.~\ref{fig:concept}(d)], by a factor of 2.7 and
independent of power from 100 to 500~mW, is not a charge-state effect in the model: a ten times
stronger channel Q lowers $S$ at 1030~nm by only 9\% and removes most of the spin contrast.
We assign it to the two-photon absorption spectrum of \NVm{}, which is resolved here with a
narrower source than in Ref.~\cite{Ji2018}.

\paragraph*{Ground-state threshold and the quartet energy.}
To obtain $\mathrm{IP}(\tA)$ we scanned 910--1000~nm at 40~mW
[Fig.~\ref{fig:threshold}(a)]. At this power the charge cycling by two-photon excitation plus
one-photon ionization, which scales with one more power of the intensity, is suppressed relative
to the two-photon steps, so the edge is not masked by it. The contrast collapses at 940--960~nm to
$C\leq0.09$ of its value at 980--1000~nm, while $S$ keeps about one third. The contrast-loss
values alone do not fix the threshold, which they trade against the strength of channel A; the
edge does. Fitted jointly with them, it gives $\mathrm{IP}(\tA)=2.63\pm0.03\eV$, with $E_q$ unchanged,
consistent with the photocurrent thresholds of 2.66(4) and 2.69(3)~eV~\cite{Bourgeois2017}. The edge could
alternatively be read as the saturation of channel A, with its threshold well below the edge
($\mathrm{IP}=2.54\eV$). The best fit of this reading [dashed lines in Fig.~\ref{fig:threshold}(a)] keeps the
contrast below 930~nm at 0.32 of its value at 980--1000~nm, whereas the measured contrast falls to
0.06--0.09; this reading is therefore excluded~\cite{SM}.

Equation~(\ref{eq:cycle}) then places the quartet at
\begin{equation}
E(\qA)-E(\dE)=0.54\pm0.04\eV
\end{equation}
[Fig.~\ref{fig:threshold}(b)], the uncertainties of $E_q$ ($\pm0.02\eV$) and of
$\mathrm{IP}(\tA)$ ($\pm0.03\eV$) being independent. Both thresholds come from the same spot, the same setup and the
same temperature, so the temperature shift of the band edge enters both onsets alike, and both are zero-phonon
thresholds. The value is consistent with the room-temperature bound
of Ref.~\cite{Shandilya2024}: that bound, $\leq0.58\eV$, assumed $\mathrm{IP}(\tA)=2.65\eV$, and
with our IP the same argument gives $\leq0.60\eV$. It is above the $\approx0.4\eV$ estimate from
separate photoionization experiments and within the calculated range (0.48~eV, HSE06, and 0.58~eV, PBE~\cite{ThieringGali2024};
0.68~eV, configuration interaction~\cite{Pfaffle2021}).

\paragraph*{Optimum excitation.}
The window of the highest figure of merit, 1060--1075~nm, lies just beyond the broadened onset of
channel Q and before the two-photon absorption falls; at 1060--1070~nm the contrast loss at equal $S$
is $-1.0$ to $+1.8\%$. The figure of merit at 1075~nm grows as $P^{0.65}$ over 88--300~mW~\cite{SM}. At 200~mW,
1075~nm exceeds 1030~nm by 46\%. For a shot-noise-limited measurement, the gain of 26--37\% over
1030--1040~nm, the wavelengths of Yb sources used so far~\cite{Nguyen2026}, shortens the
acquisition by a factor of 1.6--1.9 at equal power or, by the measured power law, allows 30--40\%
less power at equal sensitivity, while the water absorption is nearly the same (0.16 and
0.18~cm$^{-1}$ at 1040 and 1070~nm~\cite{HaleQuerry1973}). As a consequence, both the excitation and the
637--800~nm emission lie inside the near-infrared transparency window of tissue.

\paragraph*{Conclusion.}
Two-photon ODMR with a tunable source resolves the threshold of photoionization from the \NVm{}
excited state into the \NVn{} quartet, $E_q=1.23\pm0.02\eV$, and, with the ground-state
threshold from the same setup, the quartet energy $0.54\pm0.04\eV$. The spin contrast also
requires ionization of $\tE$ into the \NVn{} ground state, which sets the loss beyond 1080~nm.
The quartet threshold and the excitation spectrum together fix the optimum excitation at
1060--1075~nm, a recipe that applies to two-photon sensing with nanodiamonds in biological media.

\paragraph*{Data availability.} The data and the code of the rate model (GNU GPL v3.0) are openly
available in the HUN-REN research data repository at \href{https://doi.org/10.5158/ARP/T7NV6O}{doi:10.5158/ARP/T7NV6O}.

\begin{acknowledgments}
This work was supported by the EUREKA project quNV2.0 (2021-17465/NP/BILAT\_HU\_DE\_1, 2023--2026) via the
National Research, Development and Innovation Office (NKFIH) under Grant No.\ 2020-1.2.3-EUREKA-2022-00022.
A.G.\ acknowledges funding from the European Commission through the QuSPARC (Grant No.\ 101186889) and SPINUS
(Grant No.\ 101135699) projects. We used the AI assistant Claude
(Anthropic; Opus 4.8, 5 and 5.5, Fable 5 and 5.1) under the authors' direction to write the code for the simulations, fits and figures, to structure and
analyze the data, and to contribute to drafting and editing the text~\cite{SM}. The authors specified the model and the analysis, checked all
equations, inspected all fits and figures, and take full responsibility for the content.
\end{acknowledgments}

\clearpage
\newcount\smpages\newcount\smpage
\AddToHook{enddocument/afterlastpage}{\begingroup\hoffset=-1in\voffset=-1in
  \pdfximage{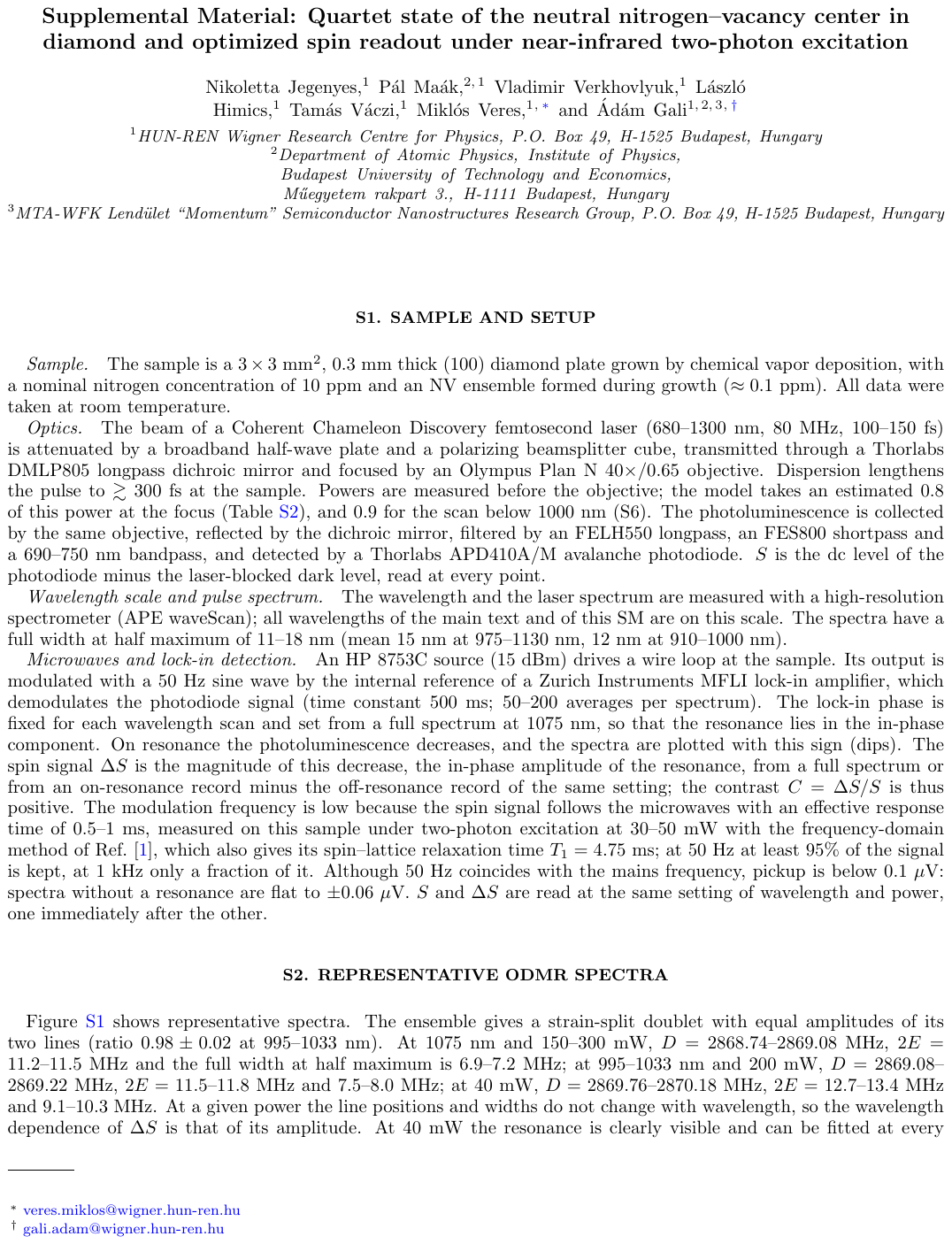}\smpages=\pdflastximagepages\smpage=0
  \loop\ifnum\smpage<\smpages\advance\smpage by 1
    \shipout\hbox{\includegraphics[page=\the\smpage]{supplement.pdf}}
  \repeat\endgroup}
\end{document}